\documentclass[10pt,twocolumn,amsmath,amssymb,aps,physrev]{revtex4-2}
\usepackage{amsmath}
\usepackage{multirow}
\usepackage{graphicx}
\usepackage{dcolumn}
\usepackage{bm}

\begin{document}

\preprint{APS/123-QE3}
\title{Kappa Entropy and its Thermodynamic Connection} 
\author{J. A. S. Lima$^{1}$} \email{jas.lima@iag.usp.br} 
\author{M. H. Benetti$^{1}$} \email{mhbenetti@usp.br}

\affiliation{$^{1}$Departamento de Astronomia (IAGUSP), Universidade de S\~ao Paulo, 05508-090 S\~ao Paulo SP,  Brasil}

\date{\today}

\begin{abstract}
   \noindent Adopting a \textit{`bottom-up'} perspective, we propose a novel two-parametric nonadditive entropy, $S_{\kappa\ell}$, associated with a Kappa-type power-law velocity distribution, $F_{\kappa\ell}(v)$, recently derived in the literature. By formulating an extended `Neo-Boltzmannian' microstate counting procedure and employing standard averaging techniques, we demonstrate that the fundamental laws of thermodynamics are preserved within this generalized power-law framework only whether $\ell=-5/2$, regardless of the values assumed by the $\kappa$-parameter.     
 \end{abstract}

\maketitle

\section{Introduction} The nonrelativistic fat-tailed  Kappa Velocity Distribution (\textbf{KVD}) was empirically proposed to explain the observed kinetic distribution of suprathermal electrons in the Earth's magnetosphere and solar winds\cite{Olbert1968,Vasyliunas1968}. Originally, it was described by the family of one-parametric power-laws:
\begin{equation}\label{E1}
 F_{\kappa} (v) = \text{A} \left(1 + \frac{v^{2}}{\kappa\omega_0^{2}}\right)^{-\kappa-1}, 
\end{equation}
where A is a normalization constant, $\omega_0$ is a thermal speed and $\kappa$ a free parameter to be empirically adjusted. In the limit $\kappa \rightarrow \infty$ the above expression reduces to the Gaussian form theoretically deduced by Maxwell from first principles\cite{Maxwell1860}. 

Later, the phenomenological approach outlined above evolved along multiple lines. The presence of fat-tailed electron distributions has been routinely confirmed in situ in a variety of space environments and observational missions \cite{Pierrard2010,Lazar2022,Benetti2023}. The family of \textbf{KVD} underwent further theoretical refinements. For instance, by replacing the exponent $-\kappa -1$ by $-\kappa$ and also by $-\kappa - 5/2$, alternative power-law forms were proposed based on distinct arguments and observational results.    
More recently, based on a fully kinetic approach, it was also demonstrated that the \textbf{KVD} in (\ref{E1}) is part of a larger set of two-parametric fat-tailed $\kappa$-distributions, collectively expressed as\cite{Lima2025b}: 
\begin{equation}\label{E2}\small
     F_{\kappa\ell}(v) = n\left(\frac{m}{2\pi k_B T }\right)^{\frac{3}{2}} \frac{ \Gamma\left(\kappa - \ell\right)}{\kappa^{\frac{3}{2}} \Gamma\left(\kappa - \ell - \frac{3}{2}\right)} \left[1 + \frac{ m v^2}{2\kappa k_B T}\right]^{-(\kappa - \ell)},
\end{equation}
where $\ell \neq \kappa$ may assume only finite values so that in the limit $\kappa \rightarrow \infty$ the Maxwellian case is also recovered. It has been found that for $\kappa = (1-q)^{-1}$ and $\ell = 0$, the above expression also reduces to the Tsallis fat-tailed power-law distribution\cite{Tsallis1988,Hasegawa1985,Silva1998,LSP2001,Leubner2002,Lima2025a}, while for arbitrary values of $\kappa$ and $\ell =-1$, it recovers the original \textbf{KVD} in, (\ref{E1}), but now with the value of $\omega_0^{2} = 2k_B T/m$ properly identified. The physical temperature $T$ has also been kinetically defined for the entire family $F_{\kappa\,\ell}(v)$, irrespective of the values of the free parameters ($\kappa,\ell$). We emphasize that the fat-tailed, power-laws (\ref{E2}) were deduced by extending Maxwell's original kinetic approach for $N$ particles confined in a volume $V$, in thermal equilibrium at temperature $T$. Interestingly, the development of Kinetic theory and statistical mechanics—from Maxwell to Boltzmann—followed a \textit{bottom-up} approach, a well-paved road that helped elucidate the obscure concept of entropy.

We advocate here that $F_{\kappa\,\ell}(v)$ is the missing link to an extended \textit{`bottom-up'} nonadditive entropic framework for power-laws and, as such, it may help clarify some conceptual difficulties underlying the widely adopted entropic (nonadditive) \textit{`top-down'} view. Naturally, the existence of two distinct levels requires their careful integration to prevent conceptual ambiguities. The interplay between $F_{\kappa\,\ell}(v)$, and $S_{\kappa\,\ell}$ can also be combined within an extended Boltzmann's approximate counting of accessible microstates. The resulting structure allows one to address unsolved issues in nonadditive statistical mechanics, including: \textit{(i)} the condition for thermal equilibrium and, more broadly, the validity of the fundamental laws of thermodynamics, and \textit{(ii)} the appropriate procedure for calculating mean quantities—whether through standard averaging or a deformed $(\kappa,\ell)$-averaging scheme. 

\begin{figure*}[t!] 
    \centering
    \includegraphics[width=\textwidth]{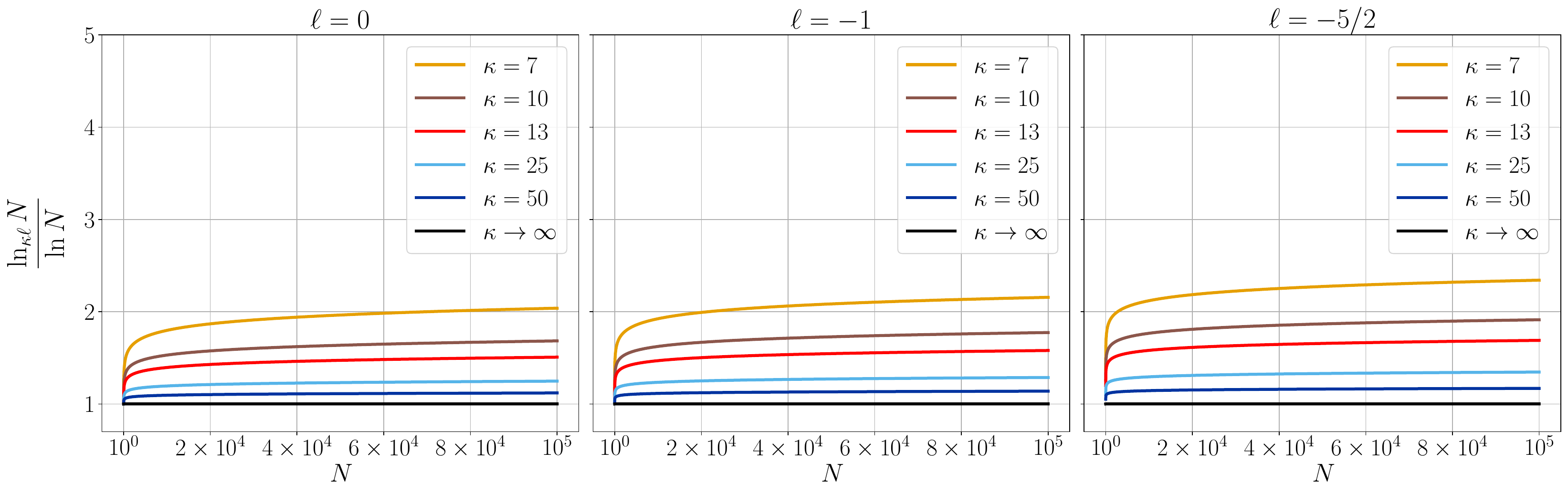}  
    \caption{Ratio between deformed and natural logarithms for different values of the parameter $ \kappa $ and some selected values of the $\ell$ parameter ($\ell = 0, -1, \, \text{and} \, -5/2 $). Black line in the bottom corresponds to the Boltzmannian limit $ \kappa \to \infty $. Note that for different values of the deformation parameter $\kappa$, the resulting curves remain nearly parallel to the standard Boltzmannian case. This means that  the deformed logarithm may likewise be approximated by a mildly corrected step-like function, analogous to the behaviour of the natural logarithm of Boltzmann' approach [cf. equations (\ref{E8})-(\ref{E10})].}
    \label{fig1}
\end{figure*}
\section{ Deformed Entropy}
On entropic grounds, the most straightforward approach to incorporate correlations underlying the power-law (\ref{E2}) is by employing the  $(\kappa,\ell)$-deformed logarithmic together with its corresponding inverse function, the deformed exponential form:
{\small
\begin{equation}\label{E3}
\ln_{\kappa\ell}(f) = \kappa\left( 1 - f^{-\frac{1}{\kappa - \ell}} \right), \,\,\,
e_{\kappa\ell}(f) = \left( 1 - \frac{f}{\kappa} \right)^{-(\kappa - \ell)}.
\end{equation}}
\noindent As expected in the limit $\kappa \to \infty$ (with $\ell \ne \kappa$), both functions recover their classical forms, $\ln_{\kappa\ell}(f) \to \ln(f)$ and $e_{\kappa\ell}(f) \to \exp(f)$. Now, to incorporate statistical correlations at the entropic level, let us define the entropy: 
\begin{equation}\label{E4}
\frac{S_{\kappa\ell}}{k_B} = \ln_{\kappa\ell} {W} = - \ln_{\kappa\ell} \frac{1}{W} =  -\ln_{\kappa\ell} \left( \frac{\prod_{i=1}^K n_i!}{N!}\right) ,
\end{equation}
where $W$ denotes the number of possible microscopic arrangements compatible with a given macroscopic configuration for distinguishable classical particles, $n_i$ is the occupation number of the $i$-th accessible microstate (or $K$-cell), and $N$ is the total number of particles. From (\ref{E4}), the entropy satisfies the following nonadditive composition rule for two independent subsystems $A$ and $B$:
\begin{equation}\label{E5}
S_{\kappa\ell}(A + B) = S_{\kappa\ell}(A) + S_{\kappa\ell}(B) + \frac{1}{\kappa k_B} S_{\kappa\ell}(A) S_{\kappa\ell}(B),
\end{equation}
and for any finite value of $\ell$ and $\kappa \to \infty$, this expression recovers the additive property of the Boltzmann entropy: 
\begin{equation}\label{E6}
    S_{B}(A+B) = S_{B}(A) + S_{B}(B),
    \end{equation}
and from (\ref{E4}) one also obtains the standard relations:
\begin{equation}\label{E7}
    \frac{S_{B}}{k_B} =\ln \left( \frac{N!}{\prod_{i=1}^K n_i!}\right) = \ln N! - \sum_i^K\ln n_i! = \ln W .  
\end{equation}
For our forthcoming purpose, rather than using the Stirling approximation to expand the factorial above, it is more convenient to adopt the step-like approximation:
\begin{equation}\label{E8}
\ln N! \approx \int_1^N \ln \chi\, d\chi = N\ln\left(\frac{N}{e}\right) + 1,
\end{equation} 
\noindent which can also be extended to each $n_i$, under the proviso that a single state overwhelmingly dominates the factorials, since in this case the small numbers are relatively negligible. Now, considering that N is constant, that is, $\sum_{i=1}^{K} n_i = N$, it follows that:
\begin{equation}\label{E9}
\frac{S_{B}}{k_B} \approx N\ln N - \sum_{i=1}^K n_i\ln n_i.
\end{equation}
This expression played a crucial role in Boltzmann's work in determining the values of $n_i$ for which $W$ reaches a maximum, thereby establishing the connection between entropy and thermodynamics. In analogy with the previous extension of Maxwell's work \cite{Lima2025b}, let us develop Boltzmann's framework by advancing to the next stage of the \textit{‘bottom-up’} kinetic approach.

 {\bf Figure  \ref{fig1}} shows the ratio between the deformed and natural logarithmic curves as a function of $\kappa$, for the values of $\ell$ usually adopted in the literature. The black line at the bottom is the limit ${\kappa \to \infty}$ when the ratio becomes unity. Notice that although slightly shifted, the deformed logarithm closely mirrors the variation of the natural logarithm. This finding suggests that some properties involving $\ln N$  can be extended to $\ln_{\kappa\ell} N$ by incorporating a slowly varying ($\kappa,\ell$)-dependent correction factor, say, $B(\kappa,\ell)$, in such a way that basic Boltzmann's results (\ref{E8}) - (\ref{E9}) might be fully recovered in the proper limit.  Hence, keeping this in mind, we write:
{\small
\begin{equation}\label{E10}
\ln_{\kappa\ell} N! \approx B\int_1^N \ln_{\kappa\ell}\chi\, d\chi = B\left[N\ln_{\kappa\ell}\left(\frac{N}{C}\right) - \ln_{\kappa\ell}\left(\frac{1}{C}\right)\right],
\end{equation}}
where $C \equiv \left[1 - 1/(\kappa - \ell)\right]^{-(\kappa - \ell)}$. Notice that $C \to e$ when $\kappa \to \infty$ so that (\ref{E10}) reduces to the classical step-like result in (\ref{E8}) assuming that $B$ also goes to unity. Now, in order to simplify the calculations for the deformed entropy, consider the following identity: $\ln_{\kappa\ell} \left(x/y \right)
=
y^{ \frac{1}{\kappa - \ell}} \left[
\ln_{\kappa\ell}(x) - \ln_{\kappa\ell}(y)
\right]$, which leads to
\begin{equation}\label{E11}
\frac{S_{\kappa\ell}}{k_B} \approx \left(1 - \frac{D}{\kappa}\right)\left[ N \ln_{\kappa\ell}(N) - \sum_i n_i \ln_{\kappa\ell}(n_i) \right], 
\end{equation}
this equation should be compared with the Boltzmannian result (\ref{E9}). We stress that the multiplicative factor $1 - D/\kappa = B(CN!)^{1/(\kappa - \ell)}$ ensures that the deformed entropy recovers the Boltzmann result in the extensive limit, as long as $\kappa \to \infty$, we have $1 - D/\kappa \to 1$. Now, given that $B$ is free within a limiting constraint, what remains is to obtain the constant value of $D$, thereby guaranteeing a consistent link between nonadditive entropy and the laws of thermodynamics (see section 4). 

\section{Equilibrium Distribution Function} Let us now determine the set of occupation numbers $ \{n_i\} $ that maximizes the deformed entropy by applying the standard variational principle. The corresponding Lagrange multipliers, $\alpha$ and $\beta$, are associated with the following constraints: 
\begin{equation}\label{E12}
\sum_{i=1}^{K} n_i = N, \qquad \sum_{i=1}^K n_i \varepsilon_i = U,
\end{equation}
respectively. Introducing these multipliers into the functional and performing the variation we have:
\begin{equation}\label{E13}
\frac{1}{k_B}\delta\left( -\ln_{\kappa\ell} \frac{1}{W} \right) - \alpha \sum_i \delta n_i - \beta \sum_i \varepsilon_i  \delta n_i = 0.
\end{equation}
By using (\ref{E11}) and taking the variation with respect to $n_i$, the extreme condition requires:
\begin{equation}\label{E14}
 \left(1 - \frac{D}{\kappa}\right)\ln_{\kappa\ell}  \left( n_i\right) = -\alpha - \beta\varepsilon_i .
\end{equation}
Note that, independently of the choice of a finite $D$, the Boltzmann result ($\ln n_i = -\alpha - \beta \varepsilon_i$) is always recovered in the limit $\kappa \to \infty$. Perhaps more interesting, the energy scale can be made to depend solely on $\beta$ by setting $D = \alpha$. Such a condition will play a key role in the thermodynamic connection as discussed below. Then, by computing the deformed exponential on both sides of (\ref{E14}), we obtain the expressions summarized in Table~\ref{tab1} for the distribution function $n_i$, the total number of particles $N$, the mean energy $U$ and the deformed partition function $Z_{\kappa\ell}$, with their respective Boltzmann's results. 

At this point, one may ask how the discrete results in Table 1 can be formulated through a smooth distribution function. This kind of transition  was also discussed by Boltzmann's much before the advent of quantum mechanics\cite{Boltzmann1877}. The discrete energy levels were assumed to be separated by an infinitesimally small quantity $\delta\varepsilon$, and the solutions $n_i $ rewritten as $ F(\varepsilon)$, a smooth density distribution in terms of the energy. 

Analogously, our variational solution derived from the deformed entropy, leads to a continuous distribution $F_{\kappa\ell}(\varepsilon) = [e_{\kappa\ell}(\alpha)]^{-1} e_{\kappa\ell}(-\beta \varepsilon)$. By identifying $\varepsilon = mv^2/2$ and normalizing the resulting $F_{\kappa\ell}(v)$ to the number of particles, the spatial homogeneity leads to the deformed velocity distribution given by (\ref{E2}). Notice that the \textbf{KVD} family, $F_{\kappa\ell}$, was originally obtained via an extended Maxwellian formulation~\cite{Lima2025b}, developed independently of the nonadditive entropic method adopted here.\\

\section{Deformed Entropy and Thermodynamics}

Let us now discuss the connections between the present deformed statistical formulation and the laws of thermodynamics. For a while, suppose constant volume ($dV = 0$). So, by inserting (\ref{E14}) into (\ref{E11}) with $D = \alpha$, while taking into account the constraints given by (\ref{E12}), the maximum combinatorial entropy becomes
\begin{equation}\label{E15}
\frac{S_{\kappa\ell}}{k_B} =
\left(1 -\frac{\alpha}{\kappa}\right) N \ln_{\kappa\ell}(N) + N\alpha + U\beta. 
\end{equation}
Now, by taking the differential and using the algebraic relation: $1 - \frac{\ln_{\kappa\ell}(N)}{\kappa} = N^{-\frac{1}{\kappa - \ell}}$  from (\ref{E3}), we find
\begin{equation}\label{E16}
\frac{dS_{\kappa\ell}}{k_B} =
N^{1-\frac{1}{\kappa -\ell}} \, d\alpha + U \, d\beta + \beta \, dU,  
\end{equation}
which can be simplified as follows: 
\begin{equation}\label{E17}
N^{-\frac{1}{\kappa - \ell}}\frac{d\alpha}{d\beta}
=     
-\frac{
\sum_i
\varepsilon_i e_{\kappa\ell}(-\beta \varepsilon_i)^{1+\frac{1}{\kappa - \ell}}
}{
\left[\sum_i 
e_{\kappa\ell}(-\beta \varepsilon_i)\right]^{1+\frac{1}{\kappa - \ell}}
} = \frac{d\ln_{\kappa\ell} Z_{\kappa\ell}}{d\beta}. 
\end{equation}
This expression was obtained by taking the natural logarithm of the normalization condition on \textit{l.h.s.} of Table \ref{tab1}, and differentiating the resulting equation. It is also known that Boltzmann's partition function satisfies: $(d\ln Z_{(B)}/d\beta) /(-U_{(B)}/N_{(B)}) = 1 $. However, in the power-law context, it is not generically obtained unless the definition of statistical averages is modified. So, many authors have introduced escort type-averages in nonadditive entropies,  e.g., $ U_q = \sum_i n_i^q \varepsilon_i$\cite{Tsallis1998}.
\begin{table}[t!]
\caption{\label{tab1} Comparing equilibrium deformed quantities and Boltzmann results. The standard Boltzmann (B) are obtained in the limit $\kappa \to \infty$ [see discussion below (\ref{E14})].}
\begin{ruledtabular}
\begin{tabular}{ll}
Deformed $(\kappa,\ell)$ formulation & (B) limit $(\kappa \to \infty)$ \\[2pt]
\hline \\[-6pt]

$\displaystyle n_i = \left[e_{\kappa\ell} \left( \alpha \right)\right]^{-1} e_{\kappa\ell}(-\beta \varepsilon_i)$ 
& $\displaystyle n_{i(B)} = e^{-\alpha} e^{-\beta \varepsilon_i}$ \\[10pt]

$\displaystyle N = \left[e_{\kappa\ell} \left( \alpha \right)\right]^{-1} \sum_i e_{\kappa\ell}(-\beta \varepsilon_i)$ 
& $\displaystyle N_{(B)} = e^{-\alpha} \sum_i e^{-\beta \varepsilon_i}$ \\[10pt]

$\displaystyle U = \left[e_{\kappa\ell} \left( \alpha \right)\right]^{-1} \sum_i \varepsilon_i\, e_{\kappa\ell}(-\beta \varepsilon_i)$ 
& $\displaystyle U_{(B)} = e^{-\alpha} \sum_i \varepsilon_i\, e^{-\beta \varepsilon_i}$ \\[10pt]

$\displaystyle Z_{\kappa\ell} = \sum_i e_{\kappa\ell}(-\beta \varepsilon_i)$ 
& $\displaystyle Z_{(B)} = \sum_i e^{-\beta \varepsilon_i}$ \\[2pt]

\end{tabular}
\end{ruledtabular}
\end{table}

\begin{figure}[t!] 
    \centering
    \includegraphics[width=0.85\columnwidth]{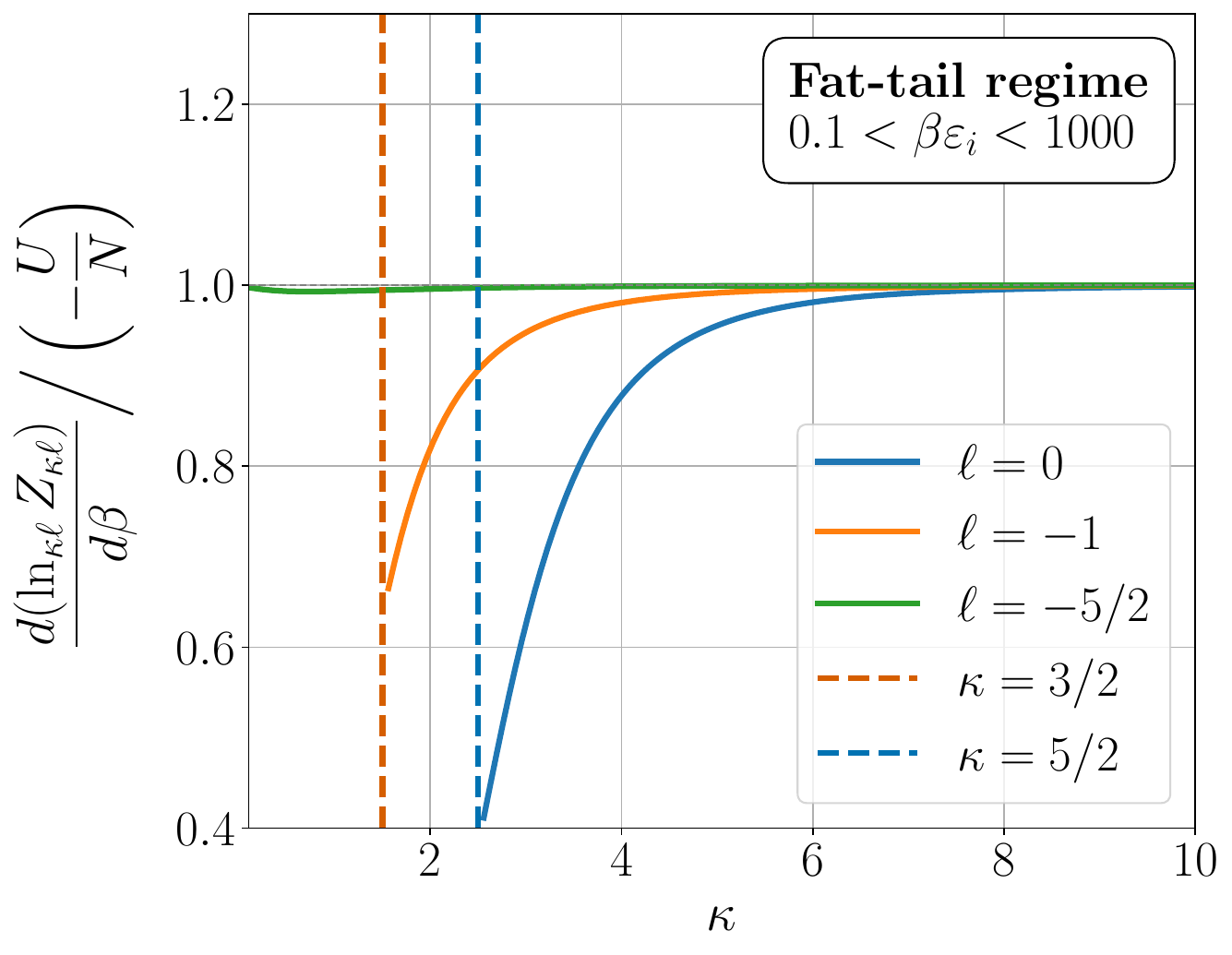}  
    \caption{
Ratio between the derivative of the deformed logarithm of the partition function and the quantity $-U/N $ for the \textbf{fat-tail} case ($ \kappa >0 $), where the distribution has a high-energy tail ($ 0.1 < \beta \varepsilon_i < 1000 $). The vertical dashed lines represent the cuts to avoid divergences in the \textbf{KVD} (\ref{E2}), for $\ell = -1 \,(\kappa > 3/2)$ and $\ell = 0 \,(\kappa > 5/2)$.
}
\label{fig2}
\end{figure}

In contrast, a different strategy is pursued in the present approach.  Constraints (\ref{E12}) show that the standard definition of averages has been applied and interesting results derived (see Table I). Nonetheless, a new line of reasoning is the following. It is known that the identity $[d\ln_{\kappa\ell}  Z_{\kappa\ell}/d\beta]/(-U/N) = 1$ is satisfied  in Boltzmann's approach ($\kappa\rightarrow \infty$). In addition, \textbf{Figure \ref{fig2}} also shows that it remains approximately valid, especially in the regime $ 0.1 < \beta \varepsilon_i < 1000 $, where the \textbf({KVD}) displays suprathermal behavior. The agreement is remarkably precise for the green curve when $ \ell = -5/2 $. This is exactly the case for which the equation of state is the same for a perfect Maxwellian gas \cite{Lima2025b}. This basic result combined with Eq.~(\ref{E17}), leads to: 
\begin{equation}\label{E18}
\frac{d\alpha}{d\beta} \approx
 -\dfrac{U}{N^{1-\frac{1}{\kappa-\ell}}} \implies  N^{1 - \frac{1}{\kappa -\ell}}\,d\alpha + U\,d\beta \approx 0.
\end{equation}
Thus, for $V= \text{constant}$ and $\ell=-5/2$, the differential entropy in (\ref{E16}) yields the standard thermodynamic relation 
\begin{equation}\label{E19}
dS_{\kappa\ell} = k_B \beta dU =\frac{dQ_{rev}}{T} = \frac{dU}{T},
\end{equation}
\noindent
where \( \beta = (k_B T)^{-1} \), as derived from the power-law formalism within the generalized Maxwellian framework [see Eq.~(\ref{E2}) and Ref.~\cite{Lima2025b}]. Moreover, the extended nonadditive entropy, \( S_{\kappa\ell} \), reinforces that the only energy scale is $k_BT$, which is essential for a consistent thermodynamic connection and also resolves recurrent controversies about the validity of the zeroth law in nonadditive statistics\cite{Nauenberg2003,Tsallis2004,Nauenberg2004,P2012,Lima2020}.

Let us now include volume variations in a simplified manner. All terms in (\ref{E16}) are dimensionless. Thus, the contribution to the variation of $S_{\kappa\ell}$ would be proportional to the fractional variation of the volume, $\Delta = dV/V$. In fact, from phase-space arguments based both in \textbf{Figure 1} and the deformed entropy formula, the contribution is $\Delta = NdV/V$, exactly the same Boltzmannian result. For the sake of generality, we consider a new contribution of the form, $dS_{\kappa_\ell} \propto \,\lambda N^{\sigma}\Delta$, where $\lambda$ and $\sigma$ are undetermined constants to be specified.  Thus, we may write 
\begin{equation}\label{E20}
N^{1 - \frac{1}{\kappa - \ell}} \, d\alpha + U\, d\beta = \lambda\, N^{\sigma} \frac{dV}{V},
\end{equation}
which reduces to (\ref{E18}) in the absence of volume changes ($dV = 0$). The differential entropy now takes the form:
\begin{equation}\label{E21}
\frac{dS_{\kappa\ell}}{k_B} = \beta\, dU + \lambda\, N^{\sigma} \frac{dV}{V}.
\end{equation}
However, as shown in the extended Maxwell approach for power laws, the equation of state $P V = N k_B T$ remains valid only in the case $\ell = -5/2$ \cite{Lima2025b}. Consequently, inserting the value of $N$ in the above expression, one may check that the standard thermodynamic relation 
\begin{equation}\label{E22}
T\, dS_{\kappa\ell} = dU + P\, dV,
\end{equation} 
is recovered for $\ell = -5/2$ since in this case $\lambda = \sigma =1$. 
\section{Final Remarks}

In this article, we  have discussed a novel two-parametric nonadditive entropy, $S_{\kappa\ell}$, associated with a Kappa-type power-law velocity distribution. It was shown that within the full class of entropies, $S_{\kappa\ell}$, aligned with the power law \textbf{KVD} only the large subclass, $S_{(\kappa,-5/2)}$, and, naturally, $F_{(\kappa,-5/2)}$, are compatible with the fundamental structure of thermodynamics. Finally, a \textit{sine qua non} condition must be emphasized: these results remain valid only when \textbf{standard averages} of the physical quantities are applied. The case involving alternative averages and their implications will be addressed in a forthcoming communication.\\   
\vspace{2pt}
\noindent\textbf{Acknowledgments:} JASL is partially supported by CNPq (310038/2019-7) and FAPESP under grants (LLAMA project, 11/51676-9 and 24/02295-2). MHB is supported by FAPESP/CNPq (24/14163-3).

\appendix


\end{document}